\documentclass[12pt]{article}

\usepackage{calphad}
\newcommand{\maps}{\qcmd{maps}}
\newcommand{\atat}{ATAT}

\newcommand{\url}[2]{\htmladdnormallink{#1}{#2}}

\newcommand{\qcmd}[1]{{\tt #1}}
\newcommand{\dqcmd}[1]{\bitem{\item[] \qcmd{#1}}}

\newcommand{\ie}{{\it i.e.}}
\newcommand{\eg}{{\it e.g.}}

\newcommand{\bitem}[1]{\begin{itemize}#1\end{itemize}}
\newcommand{\benum}[1]{\begin{enumerate}#1\end{enumerate}}

\usepackage{epsf}

\newcommand{\illuseps}[2]{
\begin{figure}[!ht]
\caption{\label{#1}#2}
\centerline{\epsfbox{#1.eps}}
\end{figure}
}

\usepackage{citesort}

\begin{document}

\vspace*{5\baselineskip}

\centerline{{\large \bf The Alloy Theoretic Automated Toolkit: A User Guide}}
\vspace{1\baselineskip}
\begin{center}
A. van de Walle,$^{a}$ M. Asta,$^{a}$ and G. Ceder$^{b}$\\
$^{a}$ Department of Materials Science and Engineering,\\
 Northwestern University, Evanston, IL 60208-3108\\
{\tt avdw@alum.mit.edu}\\
$^{b}$ Department of Materials Science and Engineering,\\
Massachusetts Institute of Technology, Cambridge, MA 02139
\end{center}

\begin{abstract}
Although the formalism that allows the calculation of alloy thermodynamic properties from
first-principles has been known for decades, its practical implementation has so far remained
a tedious process.
The Alloy Theoretic Automated Toolkit (\atat{}) drastically simplifies this procedure
by implementing decision rules based on formal statistical analysis that frees the
researchers from a constant monitoring during the calculation process and automatically
``glues'' together the input and the output of various codes, in order to provide
a high-level interface to the calculation of alloy thermodynamic properties from first-principles.
\atat{} implements the Structure Inversion Method (SIM), also known as the Connolly-Williams 
method, in combination with semi-grand-canonical Monte Carlo simulations.
In order to make this powerful toolkit available to the wide community of researchers who
could benefit from it, this article present a concise user guide outlining the steps required
to obtain thermodynamic information from {\it ab initio} calculations.
\end{abstract}

\section{Introduction}

First-principles calculations of alloy thermodynamic properties have been successfully
employed in a variety of contexts for metallic,
semi-conductor and ceramic systems, including the computation of:
composition-temperature phase diagrams, thermodynamic properties of stable
and metastable phases, short-range order in solid solutions, thermodynamic
properties of planar defects (including surfaces or antiphase and interphase
boundaries), and the morphology of precipitate microstructures
\cite{ducastelle:book,fontaine:clusapp,zunger:NATO,zunger:scord,wolverton:srorev,ceder:oxides,asta:fppheq}.

Although the formalism that allows the calculation of thermodynamic properties from
first-principles has been known for decades \cite{ducastelle:book,fontaine:clusapp,zunger:NATO}, its practical implementation remains tedious.
These practical issues limit the accuracy researchers are able to obtain 
without spending an unreasonable amount of their time writing input files for various computer codes,
monitoring their execution and processing their output.
These practical difficulties also limit the community of researchers that use these methods solely to those
that possess the necessary expertise to carry out such calculations.

The Alloy Theoretic Automated Toolkit (\atat{}) \cite{avdw:atat} drastically simplifies the practical use of these methods
by implementing decision rules based on formal statistical analysis that free the
researchers from a constant monitoring during the calculation process and automatically
``glues'' together the input and the output of various codes, in order to provide
a high-level interface to the calculation of thermodynamic properties from first-principles.
In order the make this powerful toolkit available to the wide community of researchers who
could benefit from it, this article present a concise user guide to this toolkit.

\section{Theoretical Background}

While there exist numerous methodologies that enable the calculation of thermodynamic properties
from first-principles, we will focus on the following two-step approach (see Figure
\ref{mapsemc2}).
First, a compact representation of the energetics of an alloy, known as the cluster expansion
\cite{sanchez:cexp,ducastelle:book,fontaine:clusapp,zunger:NATO}, is constructed using first-principles calculations of the formation energies
of various atomic arrangements. Second, the cluster expansion is used as a Hamiltonian
for Monte Carlo simulations \cite{newman:mc,binder:mc,dunweg:mc,laradji:mc} that can provide the thermodynamic properties of interest,
such as the free energy of a phase or short-range-order parameters as a function of temperature
and concentration. This two-step approach is essential, because the calculation of thermodynamic quantities
through Monte Carlo involves averaging the property of interest over many different atomic configurations and it would be infeasible to calculate the
energy of each of these configurations from first-principles. The cluster expansion enables the prediction of the energy of any configuration
from the knowledge of the energies of a small number of configurations (typically between 30 and 50), thus making the procedure
amenable to the use of first-principles methods.

Formally, the cluster expansion is defined by first assigning 
occupation variables $\sigma_{i}$ to each site $i$ of the \emph{parent lattice},
which is defined as the set of all the atomic sites that can be occupied by one of
a few possible atomic species. In the common case of a binary alloy system, 
the occupation variables $\sigma _{i}$ take the value $-1$ or $+1$ depending
on the type of atom occupying the site. A particular arrangement of these ``spins'' on
the parent lattice is called a \emph{configuration} and can be represented
by a vector $\mathbf{\sigma }$ containing the value of the occupation
variable for each site in the parent lattice. Although we focus here on the
case of binary alloys, this framework can be extended to arbitrary
multicomponent alloys (the appropriate formalism is presented in \cite{sanchez:cexp}).

The cluster expansion then parametrizes the energy (per atom) of the alloy as a polynomial in
the occupation variables: 
\begin{equation}
E(\mathbf{\sigma })=\sum_{\alpha }m_{\alpha }J_{\alpha }\left\langle
\prod_{i\in \alpha ^{\prime }}\sigma _{i}\right\rangle \label{eq_ce}
\end{equation}
where $\alpha $ is a cluster (a set of sites $i$). The sum is taken over all
clusters $\alpha $ that are not equivalent by a symmetry operation of the
space group of the parent lattice, while the average is taken over all
clusters $\alpha ^{\prime }$ that are equivalent to $\alpha $ by symmetry.
The coefficients $J_{\alpha }$ in this expansion embody the information
regarding the energetics of the alloy and are called the effective cluster
interaction (ECI). The \emph{multiplicities} $m_{\alpha }$ indicate the
number of clusters that are equivalent by symmetry to $\alpha $ (divided by
the number of lattice sites).

It can be shown that when \emph{all} clusters $\alpha $ are considered in
the sum, the cluster expansion is able to represent any function $E\left( 
\mathbf{\sigma }\right) $ of configuration $\mathbf{\sigma }$ by an
appropriate selection of the values of $J_{\alpha }$. However, the real
advantage of the cluster expansion is that, in practice, it is found to
converge rapidly. An accuracy that is sufficient for phase diagram
calculations can be achieved by keeping only clusters $\alpha $ that are
relatively compact (\textit{e.g.} short-range pairs or small triplets). The
unknown parameters of the cluster expansion (the ECI) can then be determined by
fitting them to the energy of a relatively small number of configurations
obtained through first-principles computations. This approach
is known as the Structure Inversion Method (SIM) or the Collony-Williams 
\cite{collwill:fit} method.

The cluster expansion thus presents an extremely concise and practical way
to model the configurational dependence of an alloy's energy. A typical well-converged cluster
expansion of the energy of an alloy consists of about 10 to 20 ECI and
necessitates the calculation of the energy of around 30 to 50 ordered
structures (see, for instance, \cite{vanderven:licoo2,gdg:linfit,ozolins:noble}).
Once the cluster expansion has been constructed, the energy of any configuration can be
calculated using Equation \ref{eq_ce} at a very small computational cost.
This enables the use of various statistical mechanical techniques such as Monte Carlo
simulations \cite{binder:mc}, the low-temperature expansion (LTE) \cite{afk:lte,ducastelle:book},
the high-temperature expansion (HTE) \cite{ducastelle:book}, or the cluster variation method (CVM)
\cite{kikuchi:cvm,ducastelle:book} to calculate thermodynamic properties and
phase diagrams. The \atat{} software implements Monte Carlo simulations, the LTE and the HTE.

\illuseps{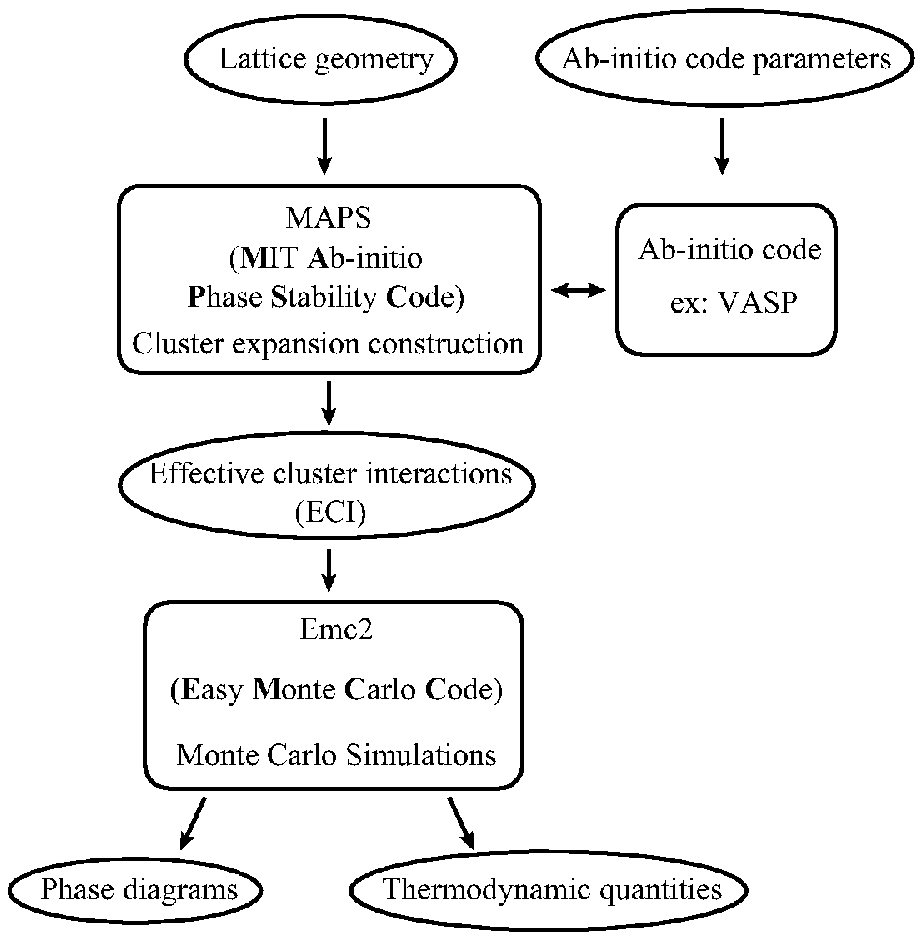}{Methodology implemented in \atat{} for the computation of thermodynamic properties from first-principles.
The automated construction of
the cluster expansion is performed by the \qcmd{maps} code. Whenever needed, \qcmd{maps} requests the calculation of
the formation energy of various atomic configurations by a first-principles code (such as \qcmd{vasp}).
The necessary input files are created and the resulting output files are parsed without requiring user intervention.
The output of \qcmd{maps} is a set of effective cluster interactions that define a computationally efficient Hamiltonian
that can be used to perform
Monte Carlo simulations with the \qcmd{emc2} code. These simulations provide thermodynamic properties and phase diagrams
that can be used to create thermodynamic databases or supplement existing ones.
}

Paralleling the two-step approach described in the previous section, \atat{} consists of two main computer programs
(see Figure \ref{mapsemc2}).
The cluster expansion construction is performed by the MIT {\it Ab initio} Phase Stability (MAPS) code \cite{avdw:maps},
while the Monte Carlo simulations are driven by the Easy Monte
Carlo Code (EMC2), developed at Northwestern University \cite{avdw:emc2}. Each of these codes will be discussed in turn.

While the present user guide describes how the \atat{} software can be used to carry out
all the steps necessary for the calculation of thermodynamic properties
from first-principles, it must be emphasized that each part of the toolkit can be used as
a stand-alone code. For instance, many users may have access to an existing cluster expansion 
obtained through the SIM or other popular methods, such as concentration-wave-based methods
(see, for instance, \cite{ducastelle:gpm,ducastelle:book,turchi:gpm}).
It is then staightforward to setup the appropriate input files to run the \qcmd{emc2} Monte Carlo code.
Alternatively, after obtaining a cluster expansion using the \qcmd{maps} code,
users could choose to calculate thermodynamic properties with the cluster variation method (CVM)
\cite{kikuchi:cvm,ducastelle:book}, as implemented in the \qcmd{IMR-CVM} code \cite{sluiter:cvmcode}.
The modularity of the toolkit actually extends below the level of the \qcmd{maps} and \qcmd{emc2} codes --- many
of the subroutines underlying these codes can be accessed through stand-alone utilities \cite{avdw:atat}.

\section{Cluster expansion construction using the MAPS code}

The \maps{} code implements the so-called Structure Inversion Method (SIM), also known as the Connolly-Williams 
method \cite{collwill:fit}. While the algorithms underlying the \maps{} code are described in \cite{avdw:maps}, the present section
focuses on its practical use.

\subsection{Input files}

The \maps{} code needs two input files: one that specifies the
geometry of the parent lattice (\qcmd{lat.in}) and one that provides
the parameters of the first-principles calculations (\qcmd{xxxx.wrap},
where \qcmd{xxxx} is the name of the first-principles code used).  The
clear separation between the thermodynamic and first-principles
calculations is a distinguishing feature of \atat{} that enables the
package to be easily interfaced with any first-principles code.  Table
\ref{exlatin} gives two annotated examples of a lattice geometry input
file.  The package includes ready-made lattice files for the common
lattice types (\eg\  bcc, fcc, hcp). It also includes an utility that
automatically constructs multiple lattice geometry input files for
common lattices. For instance,
\dqcmd{makelat Al,Ti fcc,bcc,hcp}
creates 3 subdirectories containing the appropriate input files for each specified lattice.

\begin{table}[!ht]
\caption{Examples of lattice geometry input file \qcmd{lat.in}. Typically, the coordinate system
entry is used to define the conventional unit cell so that all other entries can be
specified in the normalized coordinates that are the most natural for the symmetry
of the lattice. The input lattice parameters do not need to be exact, as the first-principles
code will optimize them.\label{exlatin}}

\hspace{0.5in}Example 1: hcp Ti-Al system

\begin{center}
\begin{tabular}{ll}
\hline
\qcmd{3.1 3.1 5.062 90 90 120       } & (Coordinate system: $a$ $b$ $c$ $\alpha$ $\beta$ $\gamma$ notation) \\
\qcmd{1 0 0                         } & (Primitive unit cell: one vector per line \\
\qcmd{0 1 0                         } & expressed in multiples of the above coordinate \\
\qcmd{0 0 1                         } & system vectors) \\
\qcmd{0         0         0   Al,Ti } & (Atoms in the lattice) \\
\qcmd{0.6666666 0.3333333 0.5 Al,Ti } \\
\hline
\end{tabular}
\end{center}

\vspace{\baselineskip}

\hspace{0.5in}Example 2: rocksalt CaO-MgO pseudobinary system
\begin{center}
\begin{tabular}{ll}
\hline
\qcmd{4.1 4.1 4.1 90 90 90} & \\
\qcmd{0   0.5 0.5} & \\
\qcmd{0.5 0 0.5} & \\
\qcmd{0.5 0.5 0} & \\
\qcmd{0 0 0 Ca,Mg  } & (``Active'' atoms in the lattice) \\
\qcmd{0.5 0.5 0.5 O} & (``spectator'' ion) \\
\hline
\end{tabular}
\end{center}
\end{table}

The first-principles input file is usually less than 10 lines long, thanks to
the dramatic improvements in the user-friendliness of most modern first-principles codes.
For instance, in the case of the widely used VASP code \cite{kresse:vasp1,kresse:vasp2}, a typical input file is given
in Table \ref{exvaspwrap}. Examples of such input files are provided with the package.
Note that \atat{} contains a utility that enables the automatic construction of
$k$-point meshes from a single parameter defining the desired target $k$-point density, the number
of $k$-point per reciprocal atom (\qcmd{KPPRA}).

\begin{table}[!ht]
\caption{Examples of first-principles code input file (example given for the \qcmd{vasp} code).
It is especially important to verify that the \qcmd{KPPRA} parameter is set sufficiently large for the system under study. \label{exvaspwrap}}

\begin{center}
\begin{tabular}{ll}
\hline
\qcmd{[INCAR]} & \\
\qcmd{PREC = high} & \\
\qcmd{ENMAX = 200} & \\
\qcmd{ISMEAR = -1} & \\
\qcmd{SIGMA = 0.1} & \\
\qcmd{NSW=41} & \\
\qcmd{IBRION = 2} & \\
\qcmd{ISIF = 3} & (See \qcmd{vasp} manual for a description of the above 6 parameters.) \\
\qcmd{KPPRA = 1000} & (Sets the $k$-point density (K Point Per Reciprocal Atom)) \\
\qcmd{DOSTATIC} & (Performs a ``static run'' --- see \qcmd{vasp} manual) \\
\hline
\end{tabular}
\end{center}

\end{table}

\subsection{Running the code}

The \maps{} code is started using the command
\dqcmd{maps -d \&}
where the option \qcmd{-d} indicates that all default values of the input parameters should be used, which is what most users
will ever need. (The optional parameters can be displayed by typing \qcmd{maps} by itself and further help is available via the command
\qcmd{maps -h}.) The trailing \qcmd{\&} character cause the command
to execute in ``background'' mode. In this fashion, \maps{} can continuously be on the lookout, responding to various ``signals'', while 
the user performs other tasks. (The ongoing discussion assumes that the code is run under a UNIX
environment within a shell such as \qcmd{sh}, \qcmd{csh}, \qcmd{tcsh} or \qcmd{bash}.)

The process of constructing a cluster expansion from first-principles calculations can be summarized as follows.
\benum{
\item Determine the parameters of the first-principles code that provide the desired accuracy.
\item Let \qcmd{maps} refine the cluster expansion.
\item Decide when the cluster expansion is sufficiently accurate.
}

Typically, one calibrates the accuracy of the first-principles calculations using the ``pure''\footnote{In the case of pseudobinary alloys
with spectator ions (\eg\ the MgO-CaO system), the ``pure'' structures would correspond to the structures where the sublattice of interest
is entirely filled with a single type of atom.} structures of the alloy system of interest.
To generate the two ``pure'' structures, type
\dqcmd{touch ready}
This creates a file called \qcmd{ready} which tells \maps{} that you are ready to calculate the energy of a structure.
Within 10 seconds, \maps{} replies with
\bitem{
\item[] \qcmd{Finding best structure...}
\item[] \qcmd{done!}
}
\maps{} has just created a directory called \qcmd{0} and, within it, a file called \qcmd{str.out} that contains the geometry of one of the two ``pure'' structures.
If you type \qcmd{touch ready} once more, the other ``pure'' structure is written to \qcmd{1/str.out}.
You now need to launch the first-principles code to calculate the energy of each structure.
Type
\dqcmd{cd 0}
to go into the directory of the first structure. Assuming that your first-principles code is called \qcmd{xxxx}, type
\dqcmd{runstruct\_xxxx \&}
After this command has successfully terminated, display the energy of that structure and
go back to the initial directory
\bitem{
\item[] \qcmd{cat energy}
\item[] \qcmd{cd ..}
}
and edit the file defining the first-principles code parameters
\dqcmd{emacs xxxx.wrap \&}
so that the precision of the calculation is increased (\eg\ increase the
$k$-point density or the cut-off of the plane-wave energy). Then you rerun
the calculations to check by how much the calculated energy has changed:
\bitem{
\item[] \qcmd{cd 0}
\item[] \qcmd{runstruct\_xxxx \&}
\item[] \qcmd{cat energy} (After the calculations are completed)
\item[] \qcmd{cd ..}
}
This process is repeated until the user is satisfied with the precision of the calculation
(that is, if the energy has become insensitive to changes in the input parameters within
the desired accuracy).\footnote{A key number to keep in mind is that an error of
25meV corresponds to 300K on a temperature scale.}
A similar study should also be performed for the other ``pure'' structure (labeled structure
\qcmd{1}) and, if one is really concerned with precision, for a few structures with
intermediate concentrations.

Once the appropriate {\it ab initio} code parameters have been determined, the fully automated
process can begin. From within the directory where \maps{} was started,
type
\dqcmd{pollmach runstruct\_xxxx}
to start the job manager that will monitor the load on your local network of workstations and ask \maps{}
to generate new structures (\ie\ atomic arrangements) whenever a processor becomes available.
Note that the first time the command is run, instructions will appear on screen
that explain how to configure the job dispatching system in accordance to your local
computing environment. Once this configuration is complete, the above command should
be invoked in the background by appending a ``\qcmd{\&}'' to it.

\subsection{Output of MAPS}

While the calculations are running, you can check on the status of the best cluster
expansion obtained so far. The file \qcmd{log.out} contains a brief description of the
status of the calculations, such as the accuracy of the cluster expansion and various
warning messages. Most of the messages pertains to the accurate prediction
of the so-called ground states of the alloy system. The ground states, which are the structures that
have the lowest energy for each given concentration, are extremely important to predict accurately
because they determine which phases will appear on the phase diagram.
The four possible messages are described below.
\bitem{
\item \qcmd{Not enough known energies to fit CE.} 
Before displaying any results, \maps{} waits until enough structural energies are known
to fit a minimal cluster expansion containing only nearest-neighbor pair interactions
and test its accuracy. Thus, the first cluster expansion is typically constructed
after at least 4 structural energies have been computed (this number may vary as a function
of the symmetry of the lattice). 
\item \qcmd{Among structures of known energy, true ground states differ from fitted ground\\ states.} 
The current cluster expansion incorrectly predicts which structures have the lowest
 energy for given concentrations, among structures whose first-principles energy is known.
The code has built-in checks to avoid this.
However, in rare instances, it may be mathematically impossible to satisfy all the constraints 
that the code imposes for a cluster expansion to be acceptable.
 This problem becomes increasingly unlikely as the number of calculated structural energies increases, 
so the user should just wait until the problem fixes itself. 
\item \qcmd{Among structures of known energy, true and predicted ground states agree.} 
Opposite of the previous message. When this message is displayed, \maps{} also displays either one of
the following two messages.
\item \qcmd{New ground states of volume less or equal to} $n$ \qcmd{predicted, see predstr.out.} 
This indicates that the cluster expansion predicts that, at some concentration, there exist other structures 
that should have an energy even lower than the one of the structures whose energy has been calculated 
from first-principles. In this case, \maps{}
will investigate the matter by generating those structures and requesting that their energy be
calculated. Once again, the user should just wait for the problem to fix itself.
The predicted ground states are flagged by a \qcmd{g} in the \qcmd{predstr.out} file, so
that you can display their energy by typing \dqcmd{grep g predstr.out}
\item \qcmd{No other ground states of} $n$ \qcmd{atoms/unit cell or less exist.}
The energies of all ground states predicted by the cluster expansion have been confirmed by first-principles calculations. Because it can be computationally intensive to perform a full
ground state search when interactions extend beyond the nearest-neighbor shell \cite{ducastelle:book}, \maps{} uses a search algorithm that merely enumerates every possible
structures having $n$ atoms or less per unit cell and uses the cluster expansion to predict their energies. The upper limit $n$ increases automatically as calculations
progress.
}

The \qcmd{log.out} file also contains two other pieces of information:
\bitem{
\item \qcmd{Concentration range used for ground state checking: [}$a$,$b$\qcmd{]} This displays the user-selected range of concentration over which ground state checking is performed (which can be specified as a command-line option of the \maps{} command: 
\qcmd{-c0=}$a$ \qcmd{-c1=}$b$).
It may be useful to relax the constraints that ground states be correctly reproduced over the whole
concentration range when it is known that other parent lattices are stable in some concentration range. In this fashion, the code can focus on providing a higher accuracy in the concentration range where the user needs it.
\item \qcmd{Crossvalidation score:} $s$. This provides the predictive power of the cluster
expansion. It is analogous to the root mean square error, except that it is specifically
designed to estimate the error made in predicting the energy for structures not included
in the least-squares fit \cite{avdw:maps}. It is defined as
\[
CV=\frac{1}{n}\sum_{i=1}^{n}\left( E_{i}-\hat{E}_{\left( i\right) }\right)
^{2}
\]
where $E_{i}$ is the calculated energy of structure $i$, while $\hat{E}%
_{\left( i\right) }$ is the predicted value of the energy of structure $i$
obtained from a least-squares fit to the $\left( n-1\right) $ other
structural energies. 
}
The \maps{} code also outputs quantitative data in various output files. The simplest
way to analyze this data is by typing
\dqcmd{mapsrep}
As illustrated in Figure \ref{mapsrep}, this command displays, in turn
\bitem{
\item The \qcmd{log.out} file described earlier.
\item The formation energy of all structures whose energy is known from first-principles calculations,
as well as the predicted energy of all structures \maps{} has in memory. The convex hull of the ground states among structures of known energy is overlaid
while the new predicted ground states (if any) are marked by an ``$\times$''.
(Note that this ground state line is only meaningful if the \qcmd{log.out} file contains ``\qcmd{Among structures of known energy, true and predicted ground states agree.}'')
\item The formation energy of all structures calculated from first-principles and associated ground state line.
\item A plot of the magnitude of the Effective Cluster Interactions (ECI) as a function of the diameter of their associated cluster (defined as the
maximum distance between any two sites in the cluster). Pairs, triplets, etc. are plotted
consecutively. This plot is useful to assess the convergence of the cluster expansion.
When the magnitude of the ECI for the larger clusters has clearly decayed to a negligible value 
(relative to the nearest-neighbor pair ECI), this is indicative of a well-converged
cluster expansion.
\item A plot of the residuals of the fit (\ie\ the fitting error) for each structure. This
information is useful to locate potential problems in the first-principles calculations.
Indeed, when first-principles calculations exhibit numerical problems, this typically results in calculated energies that are poorly reproduced by the cluster expansion.
}

\illuseps{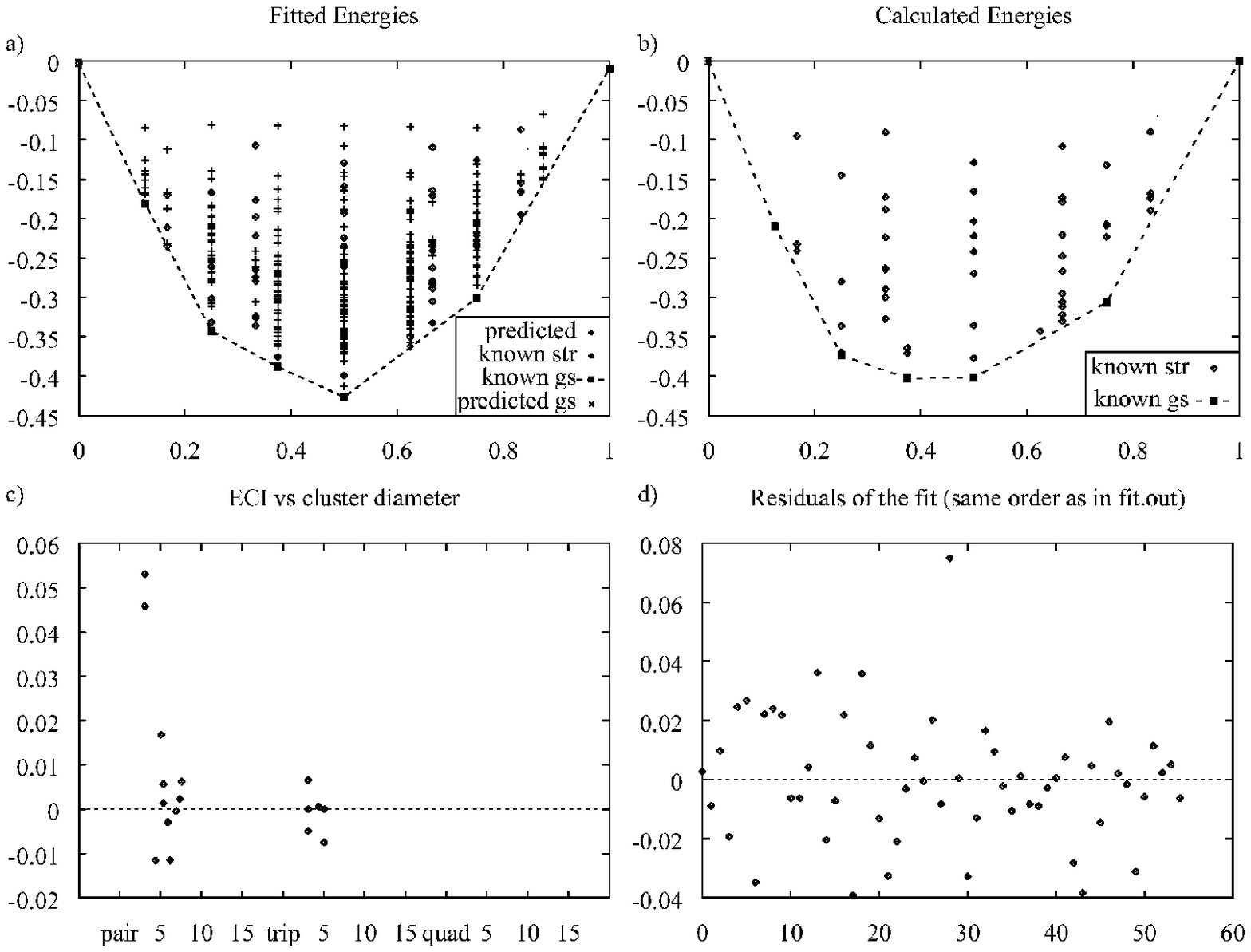}{Output of the \maps{} Code, as reported by the \qcmd{mapsrep} command.
a) Energies predicted from the cluster expansion as a function of composition for each structure generated.
``\qcmd{known str}'' denotes structures whose energy has been calculated from first-principles.
``\qcmd{known gs}'' indicate the ground states that have so far been confirmed by first-principles calculations and the dashed line
outlines the convex hull of the ground states, which serves as a threshold to detect other candidate ground states.
``\qcmd{predicted}'' denotes structures whose energy has {\em not yet} been calculated from first-principles.
``\qcmd{predicted gs}'' are structures that are predicted by the cluster expansion to be ground states, although this prediction has not yet been confirmed
by first-principles calculations.
b) Energies calculated from first-principles. ``\qcmd{known str}'' and ``\qcmd{known gs}'' are as in a), except that the energy calculated from first-principles is reported.
c) Effective Cluster Interaction (ECI) as a function of the diameter of the associated cluster and as a function of the number of sites in the cluster (\ie\ pair, triplet, etc.).
d) Residuals of the fit, that is, the difference between the first-principles energies and the energies predicted from the cluster expansion. (The abscissa refers
to the line number within the output file \qcmd{fit.out} listing all the structures with known energies.)}

When the user is satisfied with the results (which are constantly updated), 
\maps{} can be stopped by creating a file called \qcmd{stop} in the current directory
using the command:
\dqcmd{touch stop}
while the job dispatching system can be stopped by typing:
\dqcmd{touch stoppoll}
A cluster expansion can be considered satisfactory when
\benum{
\item All ground states are correctly reproduced and no new ground states are predicted.
(The \qcmd{log.out} file would then indicate that \qcmd{Among structures of known energy, true and predicted ground states agree. No other ground states of} $n$ \qcmd{atoms/unit cell or less exist.})
\item The crossvalidation score, as given in the \qcmd{log.out} file, is small (typically less than 0.025 eV).
\item Optionally, it is instructive to verify that the magnitude of the ECI decays as a function of the diameter of the corresponding cluster and as a function of the number
of sites it contains.
}

\section{Monte Carlo simulations}

The \qcmd{emc2} code implements semi-grand canonical Monte Carlo simulations \cite{newman:mc,binder:mc,dunweg:mc,laradji:mc}, where the total number
of atoms is kept fixed, while the concentration is allowed to adapt to an externally imposed difference in the chemical potential of the two types of atoms. The chemical potential
difference will be simply referred to as the ``chemical potential'' in what follows.
This ensemble offers the advantage that, for any imposed chemical potential, the equilibrium
state of the system is a single phase equilibrium, free of interfaces.\footnote{If
the simulation cell is commensurate with the unit cell of the phase under study, a requirement that the code automatically ensures.}
It also simplifies the process of calculating free energies through thermodynamic integration.
While a detailed description of the algorithm underlying this code can be found in \cite{avdw:emc2},
the current section focuses on the practical usage of the code.

\subsection{General input parameters}

The Monte Carlo code needs the following files as an input
\benum{
\item A lattice geometry file (\qcmd{lat.in}), which is the same as the input for \maps{} (see Table \ref{exlatin}).
\item Files providing the cluster expansion (the clusters used are listed in the \qcmd{clusters.out} file while the corresponding ECI are
listed in the \qcmd{eci.out} file.). These files are automatically
generated by \maps{}, although users can supply their own cluster expansion, if desired. A description of the
format of these files is available by typing \qcmd{maps -h}.
\item A list of ground states (\qcmd{gs\_str.out}), which merely provide convenient starting configurations for the simulations.
\maps{} also automatically creates this file.
}

The parameters controlling the simulation are specified as command-line options. 
The first input parameter(s) needed by the code are the phase(s) whose thermodynamic
properties are to be determined. There are two ways to invoke the Monte Carlo
simulation code. When the command \qcmd{emc2} is used, a single Monte Carlo simulation
is run to allow the calculation of thermodynamic properties of a single phase for
the whole region of chemical potential and temperature where that phase is stable.
The phase of interest is specified by a command-line option of the form
\bitem{\item[] \qcmd{-gs=}$n$,}
where $n$ is a number between $-1$ and $G-1$ (inclusive), where $G$ is the number of ground states.
The value $-1$ indicates the disordered phase while values ranging from $0$ to $G-1$ indicate the phases associated
with each ground states ($0$ denoting the ground state with the smallest composition).
When the command \qcmd{phb} is used, two Monte Carlo simulations are run simultaneously
to enable the determination of the temperature-composition phase boundary associated with
a given two-phase equilibrium. The two phases are specified by
\bitem{\item[] \qcmd{-gs1=}$n_1$ \qcmd{-gs2=}$n_2$.}
It is possible to compute a two-phase equilibrium between phases defined on a different parent lattice.
In this case, the user needs to specify the directories where the cluster expansions of
each lattice resides using the options of the form
\bitem{\item[] \qcmd{-d1=}{\it directory 1} \qcmd{-d2=}{\it directory 2}}

The accuracy of the thermodynamic properties obtained from Monte-Carlo
simulations is determined by two parameters: The size of the simulation cell
and the duration of the simulation.

The size of the simulation cell is specified by providing the radius $r$ of a sphere
through the command-line option
\bitem{\item[] \qcmd{-er=}$r$.}
As illustrated in Figure \ref{mcparam}a, the simulation cell size will be the
smallest supercell that both contains that sphere and that is commensurate with
the unit cell of the ground state of interest. 
This way of specifying the simulation cell size ensures that the system size
is comparable along every direction, regardless of the crystal structure of
the ground state of interest. It also frees the user from manually checking the
complicated requirement of commensurability. It is important that the user check
that the simulation cell size is sufficiently large for the thermodynamic properties
of interest to be close to their infinite-system-size limiting value. This can be done
by gradually increasing the system size until the calculated quantities become
insensitive to the further increases in system size, within the desired accuracy.

\illuseps{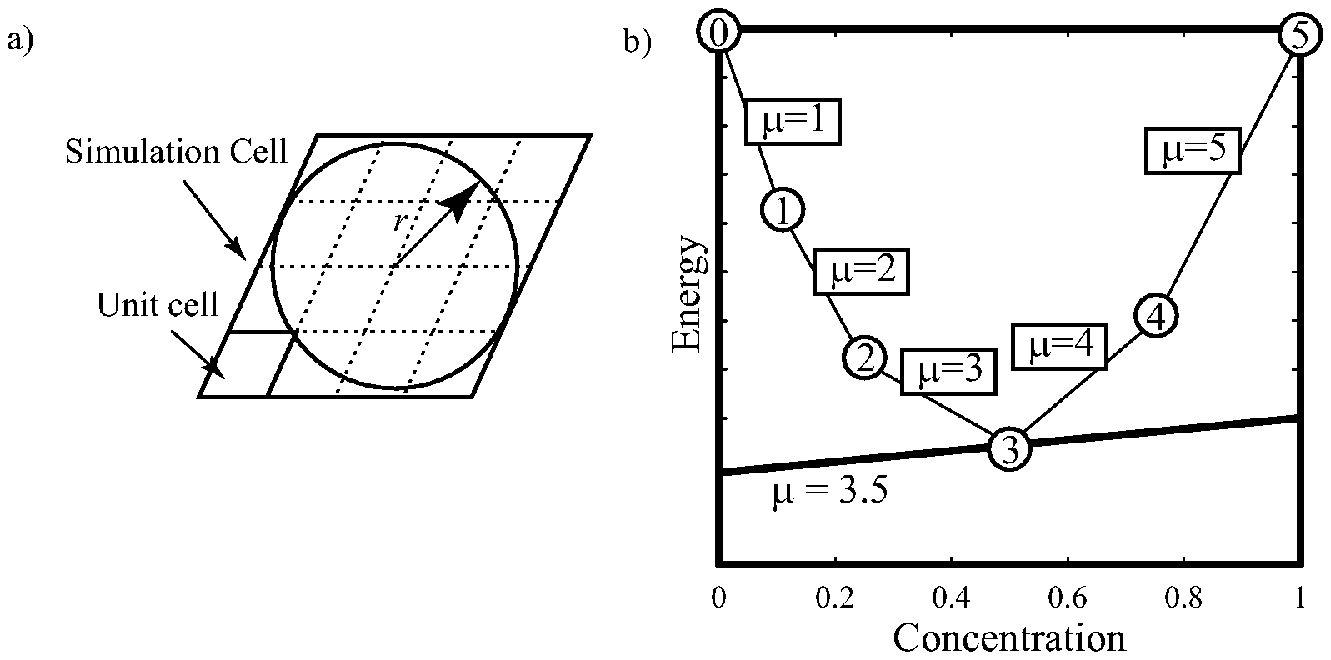}{Definitions of the quantities used to specify a) the simulation cell size and b) the chemical potential.}

The duration of the simulations is automatically determined by the code from a user-specified target precision on the atomic composition of the phase,
indicated by a command-line option of the form
\bitem{\item[] \qcmd{-dx=}$\Delta x$.}
Alternatively, the user may also manually set the number $n_{eq}$ of
Monte Carlo steps the system is allowed to equilibrate before
thermodynamic averages are computed over a certain number $n_{avg}$ of Monte Carlo steps using the options
\bitem{\item[] \qcmd{-eq=}$n_{eq}$ \qcmd{-n=}$n_{avg}$.}

The Monte Carlo code also needs additional parameters that specify which portion
of a phase's free energy surface needs to be computed. With \qcmd{emc2}, the range of
temperatures to be scanned are specified in either one of the
following two ways:
\bitem{
\item[] \qcmd{-T0=}$T_0$ \qcmd{-T1=}$T_1$ \qcmd{-dT=}$\Delta T$ (for steps in direct temperature)
\item[] or \qcmd{-T0=}$T_0$ \qcmd{-T1=}$T_1$ \qcmd{-db=}$\Delta (1/T)$ (for steps in reciprocal temperature).
}
The temperature steps in reciprocal temperature ($\Delta (1/T)$) can be useful when
calculations are started from infinite temperatures down to a finite temperature.
The \qcmd{-T1} and \qcmd{-dT} (or \qcmd{-db}) options can be omitted if calculations at a single temperature
are desired. Since the program automatically stops when a phase transition is detected,
it is not necessary to know in advance the temperature range of stability of the phase. The
user only needs to ensure that the initial temperature lies within the region of stability
of the phase of interest. An obvious starting
point is $T_0=0$, since the ground state is then stable, by definition.
With the \qcmd{phb} code, the syntax is
\bitem{
\item[] \qcmd{-T=}$T$ \qcmd{-dT=}$\Delta T$
}
If the \qcmd{-T} option is omitted, calculations start at absolute
zero.\footnote{Temperature steps in reciprocal temperature are not needed, because
a two-phase equilibrium never extents up to infinite temperature.}
The energy and temperature units used are set by specifying the Boltzman's constant with the
command-line option
\bitem{\item[] \qcmd{-k=}$k_B$.}
A value of $8.617\times 10^{-5}$ corresponds to energies
in eV and temperatures in Kelvin.

With \qcmd{emc2}, the range of chemical potentials to be scanned needs to be specified.
Once again, only the starting point really matters, because the code will stop when a
phase transition is reached.
By default, chemical potentials are given in a dimensionless form, so as to
facilitate the link between the value of the chemical potential and the
phase that it stabilizes. For instance, a chemical potential equal to $3.0$ is
such that it would stabilize a two phase equilibrium between phase number
$2$ and phase number $3$ at absolute zero (see Figure \ref{mcparam}b). A chemical potential between 
$3.0$ and $4.0$ stabilizes phase number $3$ at absolute zero. While these ranges of
stability are no longer exact at finite temperature, this dimensionless chemical
potential still provides easy-to-interpret input parameters. The syntax is
\bitem{
\item[] \qcmd{-mu0=}$\mu_0$ \qcmd{-mu1=}$\mu_1$ \qcmd{-dmu=}$\Delta \mu$
}
where $\Delta \mu$ is the chemical potential step between each new simulation.
Chemical potentials
can also be entered in absolute value (say in eV, if the energies are in eV) by
specifying the \qcmd{-abs} option.
Note that the output files always give the absolute chemical potentials, so that
thermodynamic quantities can be computed from them.
With \qcmd{phb}, the initial chemical potential is optional when starting from absolute
zero because the code can determine the required value from the ground state energies.
It can nevertheless be specified (in absolute value) with the \qcmd{-mu=}$\mu$ option, if a finite temperature starting point is desired.

A list of the command line options of either the \qcmd{emc2} or \qcmd{phb} codes can be displayed by simply typing either command by itself.
More detailed help is displayed using the \qcmd{-h} option.

\subsection{Examples}

We now give simple examples of the usage of these commands. Consider the calculation
of the free energy of the phase associated with ground state number $1$ as a function of concentration and temperature. Then, the required commands could, for instance, be
\bitem{
\item[] \qcmd{emc2 -gs=1 -mu0=1.5 -mu1=0.5 -dmu=0.04 -T0=300 -T1=5000 -dT=50 -k=8.617e-5 -dx=1e-3 -er=50 -innerT -o=mc10.out}
\item[] \qcmd{emc2 -gs=1 -mu0=1.5 -mu1=2.5 -dmu=0.04 -T0=300 -T1=5000 -dT=50 -k=8.617e-5 -dx=1e-3 -er=50 -innerT -o=mc12.out}
}
(The only difference in the two command lines is the value of \qcmd{-mu1} and the
output file name, specified by the \qcmd{-o} option.)
These commands separately compute the two ``halfs'' of the free energy surface, corresponding to
the values of the chemical potential below and above the ``middle'' value of $1.5$ 
which stabilizes ground state $1$ at absolute zero. 
This natural separation allows you to run each half calculation on a separate processor
and obtain the results in half the time.
The values of \qcmd{-dmu}, \qcmd{-dT}, \qcmd{-dx} and \qcmd{-er} given here are typical
values. The user should ensure that these values are such that the results are converged.
Note that, thanks to the way these precisions parameters are input, if satisfying values
have been found for one simulation, the same values will provide a comparable accuracy
for other simulations of the same system.
The option \qcmd{-innerT} indicates that the inner loop of the sequence of simulations scans the temperature axis while the outer loop scans the chemical potential.
In this fashion, the point of highest temperature in the region of stability of the phase will be known early during the calculations.
If the user is more interested in obtaining solubility limits early on, this option can be omitted and the inner loop with scan the chemical potential axis.
In either cases, the code exits the inner loop (and the outer loop, if appropriate) when it encounters a phase transition.

The \qcmd{emc2} code thus enables the automated calculation of the whole free energy surface of a given phase,
as illustrated in Figure \ref{mcout}a. Such free energy surfaces can be used as an input to construct thermodynamic databases or supplement
existing ones. To facilitate this process, a utility that converts the output of \qcmd{emc2} into input files for the fitting module of ThermoCalc is provided.

\illuseps{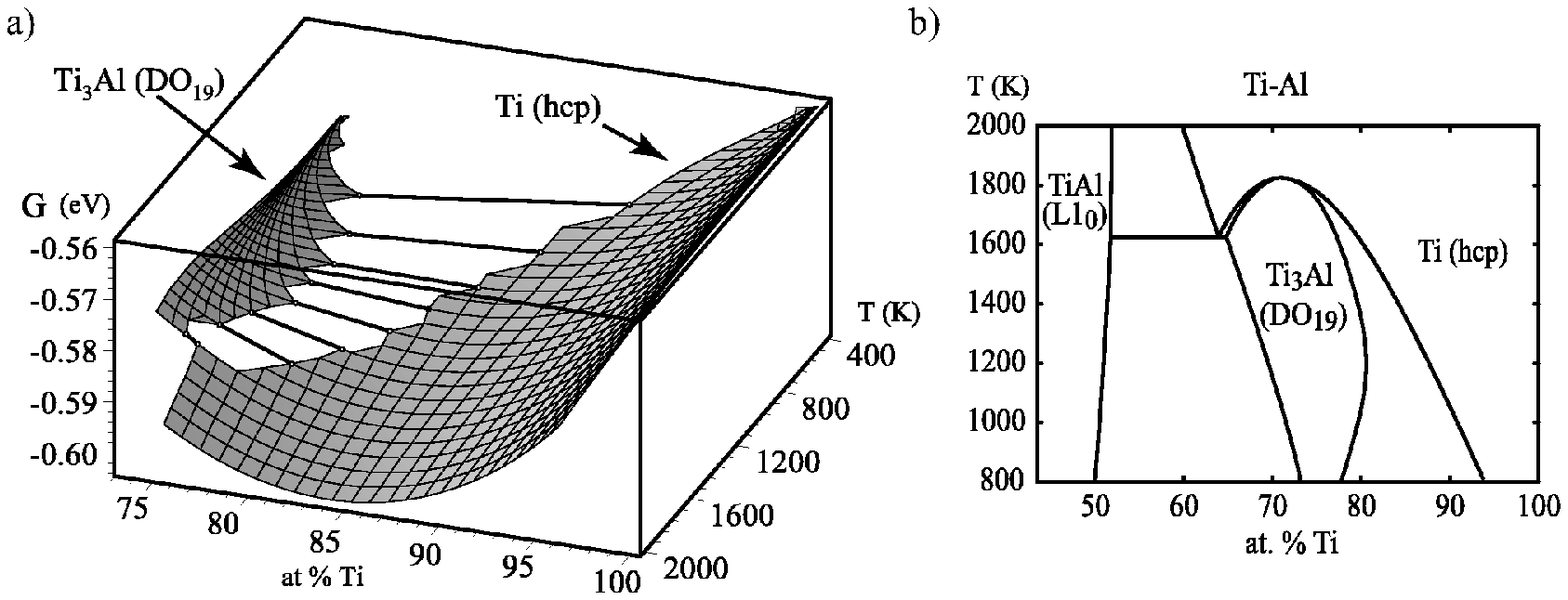}{Output of Monte Carlo codes. a) The \qcmd{emc2} provides free energy surfaces as a function of temperature $T$ and composition $x$.
(For clarity, the common tangent construction (thick lines) is drawn over the calculated free energy.)
b) The \qcmd{phb} command generates temperature-composition phase diagrams. The calculational details
underlying these results can be found in \cite{avdw:maps,avdw:emc2}}

While the above examples focus on the calculation of a phase's thermodynamic
properties over its whole region of stability, one may be interested in directly computing
the temperature-composition phase boundary without first constructing a full free energy surface. To accomplish this task, a typical command-line for the \qcmd{phb} program would be
\dqcmd{phb -gs1=0 -gs2=1 -dT=25 -dx=1e-3 -er=50 -k=8.617e-5 -o=ph01.out}
This command computes the two phase equilibrium between phase 0 and phase 1,
starting at absolute zero and incrementing temperature in steps of $25$ K.
The output file \qcmd{ph01.out} contains the temperature-composition phase boundary
of interest, as well as the chemical potential stabilizing the two-phase equilibrium
as a function of temperature. This output can be used to generate phase diagrams,
as illustrated in Figure \ref{mcout}b.

The program automatically terminates when the ``end''
of the two-phase equilibrium has been reached. If the two-phase equilibrium disappears because
of the appearance of a third phase, two new two-phase equilibria have to be separately
calculated. To do so, one uses the final temperature $T$ and chemical potential $\mu$ given in
the output file as a starting point for two new \qcmd{phb} runs:
\bitem{
\item[] \qcmd{phb -T=}$T$ \qcmd{-mu=}$\mu$\qcmd{ -gs1=0 -gs2=-1 -dT=25 -dx=1e-3 -er=50 -k=8.617e-5 -o=ph0d.out}
\item[] \qcmd{phb -T=}$T$ \qcmd{-mu=}$\mu$\qcmd{ -gs1=-1 -gs2=1 -dT=25 -dx=1e-3 -er=50 -k=8.617e-5 -o=phd1.out}
}
In the above example, it is assumed that the new phase appearing is the disordered phase (indicated by the number $-1$), which will usually be the case.
Of course, it is also possible that a given two-phase equilibrium terminates because one of the
two phases disappears. In this case, only one new calculation needs to be started, as in
the following example:
\bitem{
\item[] \qcmd{phb -T=}$T$\qcmd{-mu=}$\mu$\qcmd{ -gs1=0 -gs2=2 -dT=25 -dx=1e-3 -er=50 -k=8.617e-5 -o=phd1.out}
}
Note that phase $1$ has been replaced by phase $2$. Finally, it is also possible that the two-phase equilibrium
terminates because the concentration of each phase converges to the same value, a situation
which requires no further calculations.
The user can easily distinguish these three cases by merely comparing the
final composition of each phase.

\subsection{Interpreting the output files}

The output file of \qcmd{emc2} reports the value of all calculated thermodynamic functions for each 
value of temperature and chemical potential scanned.
The quantities reported include
\bitem{
\item Statistical averages over Monte Carlo steps, such as energy, concentration,
 short-range and long-range order parameters.\footnote{For efficiency reasons, the long range order parameters are only
 calculated when starting from an ordered phase.}
\item Integrated statistical averages, such as the Gibbs free energy $G$ or the semi-grand-canonical potential $\phi=G-\mu x$.
\item The result of common approximations, namely, the low temperature expansion (LTE) \cite{ducastelle:book,afk:lte,woodward:sgce}, the mean-field (MF) approximation
and the high temperature expansion (HTE) (see, for instance, \cite{ducastelle:book}).
}
While quantities obtained from statistical averages over Monte Carlo steps are valid for all temperatures and chemical potentials,
caution must be exercised when interpreting the result of the various approximations or when looking at the integrated quantities.
The LTE, MF and HTE approximations are only accurate in a limited range of temperature and it is the
responsibility of the user to assess this range of validity.
Also, the free energy or the semi-grand-canonical potential are obtained
from thermodynamic integration and are thus only valid if the starting point of the integration
is chosen appropriately. 
By default, the low temperature expansion value is used as a starting point whenever the
phase of interest is a ground state, while the high temperature expansion is used when
the phase of interest is the disordered state.
Hence, to obtain absolute values of the semi-grand-canonical potential, one must ensure that
the calculations are started at a sufficiently low temperature (or sufficiently high temperature,
in the case of the disordered phase). This can be checked by comparing the Monte Carlo estimates with the LTE (or HTE) estimates and verifying that they agree for the first
few steps of the thermodynamic integration. A user-specified starting point for $\phi$ (\eg\ obtained from
an earlier Monte Carlo simulation) can be indicated using the option
\bitem{\item[] \qcmd{-phi0=}$\phi$}
Note that, unlike \qcmd{emc2}, the \qcmd{phb} code automatically makes use of the low temperature expansion whenever
it is sufficiently accurate in order to save a considerable amount of computational time.

By default, the code reports the thermodynamic quantities associated
with the semi-grand-canonical ensemble, such as the
semi-grand-canonical potential $\phi$. The command-line option \qcmd{-can}, instructs the code to add $\mu x$ to all 
appropriate thermodynamic quantities, so that the code outputs the more commonly used canonical quantities, such as the
Gibbs free energy $G$ and the internal energy $E$.

\section{Future developments}

At the present time, \atat{} only handles binary systems. However,
a multicomponent version of the toolkit is under development.
About 75\% of the code has already been written with multicomponent systems
in mind.

Although the present tutorial does not discuss the topic, \atat{} also implements reciprocal space cluster expansions and, in particular,
the constituent strain formalism \cite{laks:recip}. A tutorial on the use of this feature will be
available shortly on our web site \cite{avdw:atat}.

In its current stage of development, the \atat{} package only focuses on
configurational sources of entropy. To obtain quantitative agreement with
experimental measurements (such as calorimetry), it may be necessary to explicitly add 
the vibrational and electronic contributions to the free energy \cite{avdw:vibrev}.
A simple correction that makes the calculated free energy comparable to experiments
would be to add a concentration-weighted average of the known experimental free energies 
of the appropriate elements. Of course, such a shift does not affect the calculated phase diagram,
unless it exhibits phases that are superstructures of distinct parent lattices.
In the near future, a module that computes electronic and vibrational free energy contributions
from first-principles will be added. (In fact, the Monte Carlo code already includes all the features
required to accommodate nonconfigurational entropy, such as allowing for temperature-dependent ECI.)

\section{Conclusion}

The Alloy Theoretic Automated Toolkit (\atat{}) drastically simplifies the practical implementation
of the Connolly-Williams method, in combination with semi-grand-canonical Monte Carlo simulations,
thus providing a high-level interface to the calculation of thermodynamic properties from first-principles.
This toolkit enables researcher to focus on higher-level aspects of first-principles thermodynamic calculations
by encapsulating the intricate details of the calculations under an easy-to-use interface.
It also makes these powerful methodologies readily available to the wide community of researchers who
could benefit from it.

\section*{Acknowledgements}

This work was supported by the NSF under program DMR-0080766
and by the U.S. Department of Energy, Office of Basic
Energy Sciences, under contract no. DE-F502-96ER 45571.


\end{document}